# Time Temperature Superposition in Soft Glassy Materials

Rahul Gupta, Bharat Baldewa and Yogesh M Joshi*

Department of Chemical Engineering, Indian institute of Technology Kanpur,

Kanpur 208016, India

* Corresponding author, email: joshi@iitk.ac.in

**Abstract:**

Soft glassy materials are out of thermodynamic equilibrium and show time dependent slowing down of the relaxation dynamics. Under such situation these materials follow Boltzmann superposition principle only in the effective time domain, wherein time dependent relaxation processes are scaled by a constant relaxation time. In this work we extend effective time framework to successfully demonstrate time – temperature superposition of creep and stress relaxation data of a model soft glassy system comprised of clay suspension. Such superposition is possible when average relaxation time of the material changes with time and temperature without affecting shape of the spectrum. We show that variation in relaxation time as a function of temperature facilitates prediction of long and short time rheological behavior through time – temperature superposition from the experiments carried out over experimentally accessible timescales.

## I. Introduction:

Very high viscosity soft materials such as concentrated suspensions and emulsions, colloidal gels, foams, pastes do not reach thermodynamic equilibrium over practical timescales due to structural arrest.[1-3] However the natural tendency of materials to explore phase space in order to progressively attain lower potential energy state causes evolution of structure and viscoelastic properties as a function of time.[4-7] Rheologically these materials typically demonstrate yield stress and thixotropic behavior.[8-10] Time dependency associated with these materials, however, does not allow them to obey the most fundamental linear viscoelastic theorem, the Boltzmann superposition principle in its conventional form.[11] Consequently a very powerful rheological tool: time – temperature superposition is also not applicable to this class of materials. Inapplicability of these basic linear viscoelastic principles restricts rheological modeling capacity of these industrially important soft materials. Recently our group employed effective time framework originally due to Hoffman[12] and successfully demonstrated application of Boltzmann superposition principle in the effective time domain to variety of soft glassy materials.[13, 14] In this paper we extend effective time framework to propose a procedure to carry out time – temperature superposition in the soft glassy materials.

In the limit of linear response, according to Boltzmann superposition principle, relaxation modulus (and equivalently creep compliance) depends only on time elapsed since application of deformation field.[15, 16] However, for materials that undergo time evolution of rheological properties, the relaxation modulus and creep compliance show an additional dependence on time at which the deformation field was applied.[11] Under such situation, it is customary to employ effective time approach, wherein real time with time dependent relaxation time is transformed to effective time domain where relaxation time is constant.[11, 17] Effective time scale $\xi$ is defined as:

$\xi(t)=\int_0^t \tau_0 dt'/\tau(t')$.[11, 14, 18] Since in an effective time domain relaxation time remains constant, which we consider to be an arbitrary constant $\tau_0$, Boltzmann superposition principle can be restated by replacing time elapsed since application of deformation field by effective time elapsed since application of deformation field. Therefore, in an effective time domain modified Boltzmann superposition principle can be expressed as:[11, 14, 18]

$$\sigma(t)=\int_{-\infty}^{t} G\left(\xi(t)-\xi(t_w)\right)\frac{d\gamma}{dt_w}dt_w \text{ or } \gamma(t)=\int_{-\infty}^{t} J\left(\xi(t)-\xi(t_w)\right)\frac{d\sigma}{dt_w}dt_w, \tag{1}$$

where $t$ is present time, $t_w$ is time at which deformation was applied, $G$ is relaxation modulus, $J$ is creep compliance, $\gamma$ is strain and $\sigma$ is stress. Typically for glassy materials relaxation time is observed to show power law dependence on time $\tau = A\tau_m^{1-\mu}t'^{\mu}$, where $\tau_m$ is microscopic relaxation time and $A$ is constant.[14, 17, 18] Therefore effective time elapsed since application of deformation field is given by:

$$\xi(t)-\xi(t_w)=\int_{t_w}^{t} \tau_0 dt'/\tau(t') = \frac{\tau_0\tau_m^{\mu-1}}{A}\left[\frac{t^{1-\mu}-t_w^{1-\mu}}{1-\mu}\right]. \tag{2}$$

Usually rheological behavior of any real material is dictated by distribution of relaxation times. According to definition of effective time scale, every relaxation mode will lead to different effective time scale. However if deformation field does not affect the shape of relaxation time distribution, eq. (1) can still be used to describe viscoelastic behavior of soft glassy materials as recently described by Baldewa and Joshi.[13]

In the literature effective time approach was first applied to glassy materials (amorphous polymers) by Struik.[17] However, unlike the treatment of effective time discussed in the present paper, Struik considered only the limit of $t-t_w << t_w$ (process time much smaller than aging time) in order neglect aging during the process time. In this limit eq. (2) leads to:

$\xi(t)-\xi(t_w)=\left[\tau_0\tau_m^{\mu-1}\left(t-t_w\right)\right]/\left[At_w^{\mu}\right]$ (Struik used different procedure to get similar result in the limit: $t-t_w<<t_w$. For convenience we have derived this expression directly from eq. (2)). This limiting expression suggests that creep compliance or relaxation modulus will demonstrate time – aging time superposition, if process time $\left(t-t_w\right)$ is normalized by that factor of relaxation time which depends on aging time $\left(t_w^{\mu}\right)$. Many researchers demonstrated validity of time – aging time superposition in the limit of $t-t_w<<t_w$ (also known as Struik protocol) for amorphous polymers,[17, 19] spin glasses[20, 21] and soft glassy materials.[22, 23 24-26] In order to avoid errors induced in the analysis because of approximation of eq. (2) and consideration of narrow range of process time, time – aging time superposition was attempted directly in the effective time domain and in recent years its application was successfully demonstrated for variety of soft glassy materials.[13, 14, 27, 28]

Observation of superposition in the effective time domain also leads to prediction of long time behavior of creep compliance or relaxation modulus.[13, 14] Prediction of long time behavior from time - aging time superposition is aided by change in relaxation time at different aging times. Similar to aging time, deformation field and temperature also influence the relaxation time and its dependence on aging time. Recently Baldewa and Joshi studied effect of deformation field and observed that it induces nonlinear effects thereby affecting the distribution of relaxation times.[13] Awasthi and Joshi[24] studied effect of temperature on aging and also proposed time – temperature superposition for creep flow of Laponite suspension using the Struik protocol. Therefore, the previous methodology was strongly limited by the predictive capability.[24] In this work we study the superposition behavior of creep compliance and relaxation modulus in an effective time domain at different temperatures. We demonstrate time – temperature superposition in an effective time domain and show that it leads to greater predictive capacity of long and short time rheological behavior.

## II. Material and experimental procedure

In this work we use 2.8 weight % aqueous suspension of Laponite RD,® which is known to demonstrate various characteristic features of soft glassy behavior.[29] Laponite RD consists of disk shaped particles having diameter around 30 nm and thickness 1 nm and was procured from Southern Clay Products Inc.[30] Dry Laponite powder was mixed with ultrapure water having pH 10 using Ultra Turrex drive for a period of 30 min. Subsequently the suspension was stored for a predetermined period in sealed polypropylene bottles. The detailed preparation protocol to prepare aqueous suspension of Laponite is discussed elsewhere.[31]

In this work we carry out creep and step strain experiments using Anton Paar MCR 501 rheometer (Couette geometry, bob diameter 5 mm and gap 0.2 mm). We carry out these experiments at 5 different temperatures in the range (15 to 55°C). For creep experiments we use 80 days old Laponite suspension while for step strain experiments we use 54 days old Laponite suspension. It is important to note that due to irreversible aging demonstrated by Laponite suspension over duration of days, 80 days and 54 days old Laponite suspension should be considered as different systems altogether.[31] For creep experiments a new sample is used in each experiment, while in case of step strain experiments the same sample was used for all the experiments carried out at a given temperature. At the beginning of each experiment, after the thermal equilibrium was reached, sample was shear melted by applying oscillatory flow field having strain magnitude of 1500 at frequency of 0.1 Hz for 15 min. Shear melting is necessary to obtain the same initial condition in all the experiments and therefore usage of the same sample or a new sample does not make any difference. Subsequent to shear melting, sample was allowed to age. The corresponding evolution of elastic modulus during aging was probed by applying small amplitude oscillatory shear with strain magnitude 0.5 % at frequency 0.1 Hz. After carrying out aging for a predetermined period of time, a constant shear stress (5 Pa) or step strain (3 %) was applied to the suspension.

In order to avoid drying and contamination of $CO_2$, the free surface of suspension was covered with a thin layer of low viscosity Silicone oil.

### III. Results and Discussion:

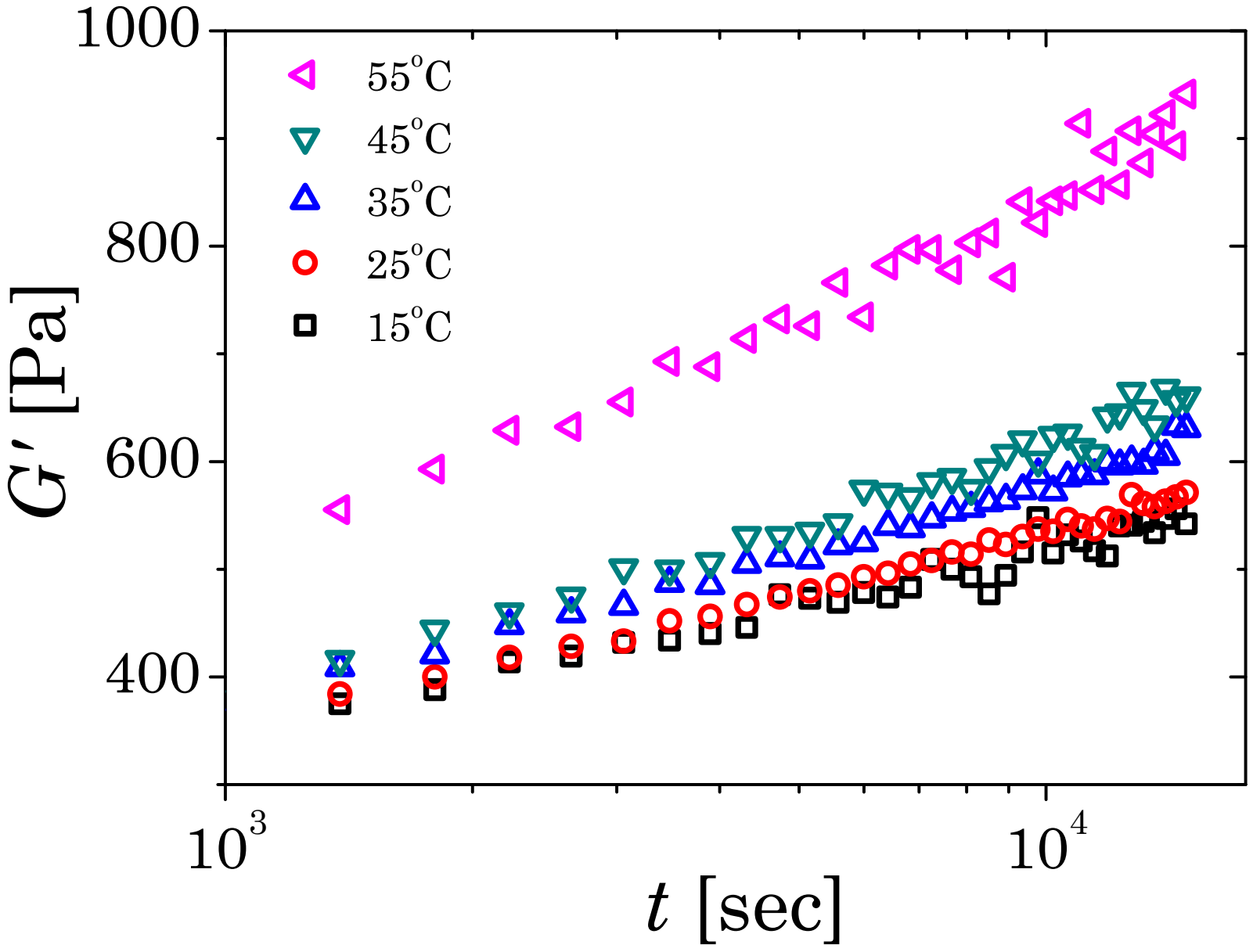

**Figure 1.** Evolution of elastic modulus as a function of time for 54 days old aqueous Laponite suspension. Evolution shifts to lower times for experiments carried out at higher temperatures.

In figure 1 evolution of elastic modulus is plotted as a function of aging time. It can be seen that evolution of elastic modulus shifts to lower aging times for experiments carried out at higher temperatures. Aging in soft glassy materials is usually accompanied by enhancement of characteristic relaxation time as well as modulus as a function of waiting time. If we consider caged entity in a glassy material to have barrier height associated with an energy well to be $E$, then relaxation time can be considered to have Arrhenius dependence on $E$ given by: $\tau = \tau_m \exp\left(E/k_B T\right)$.[11] Typically, from scaling analysis elastic modulus can be represented as energy density and can be expressed as: $G' = E/b^3$,[32] where $b$ is characteristic length-scale associated with the glassy material. Therefore enhancement of relaxation time and modulus with time suggests a state of the system with greater barrier height of the energy well (or a lower potential energy state). Shift in the evolution of $G'$ to low aging times at higher temperatures therefore suggests decrease in microscopic time (timescale associated with microscopic rearrangement) at which system explores the phase space and lowers its energy. The evolution of elastic modulus as a function of aging time then can be represented by: $G' = G_0 \ln\left(t_w/\tau_m\right)$, which can be considered to have Arrhenius dependence on temperature (and therefore becomes fast at higher temperatures). Enhanced aging at higher temperatures appears to be a characteristic feature of many glassy materials such as molecular glasses,[33] and various soft glassy materials.[34-37]

Subsequent to aging step, we perform creep experiments on 80 days old aqueous Laponite suspension samples. The corresponding evolution of creep compliance for experiments carried out at different aging times is plotted in the inset of figure 2a. It can be seen that compliance induced in the material decreases for the experiments carried out at higher aging times. We also carry out step strain experiments on 54 days old Laponite suspension. The stress relaxation subsequent to step strain, plotted in the inset of figure 2b, shows slower relaxation for experiments carried out at higher aging times. The insets in figure 2a and 2b therefore demonstrate that compliance as well as stress relaxation modulus do not depend solely on time elapsed since application of deformation field but also show an additional dependence on time at which deformation was applied. This behavior is not in agreement with Boltzmann superposition principle. However, as suggested before, Boltzmann superposition principle can be applied to such class of materials in an effective time domain, wherein it is proposed that compliance is a sole function of effective time elapsed since application of deformation field. In figure 2a and 2b we plot vertically shifted compliance and relaxation modulus respectively, as a function of $\left[t^{1-\mu}-t_w^{1-\mu}\right]/(1-\mu)$, wherein we assume power law dependence of relaxation time on aging time ($\tau = A\tau_m^{1-\mu}t_w^{\mu}$). It can be seen that compliance and stress relaxation modulus data shows an excellent superposition in the effective time domain for a certain value of $\mu$. The observed time – aging time superposition also leads to validation of Boltzmann superposition principle in the effective time domain. Importantly the fitted value of $\mu$ suggests rate of enhancement of relaxation time as a function of aging time: $\mu = d\ln\tau/d\ln t_w$.

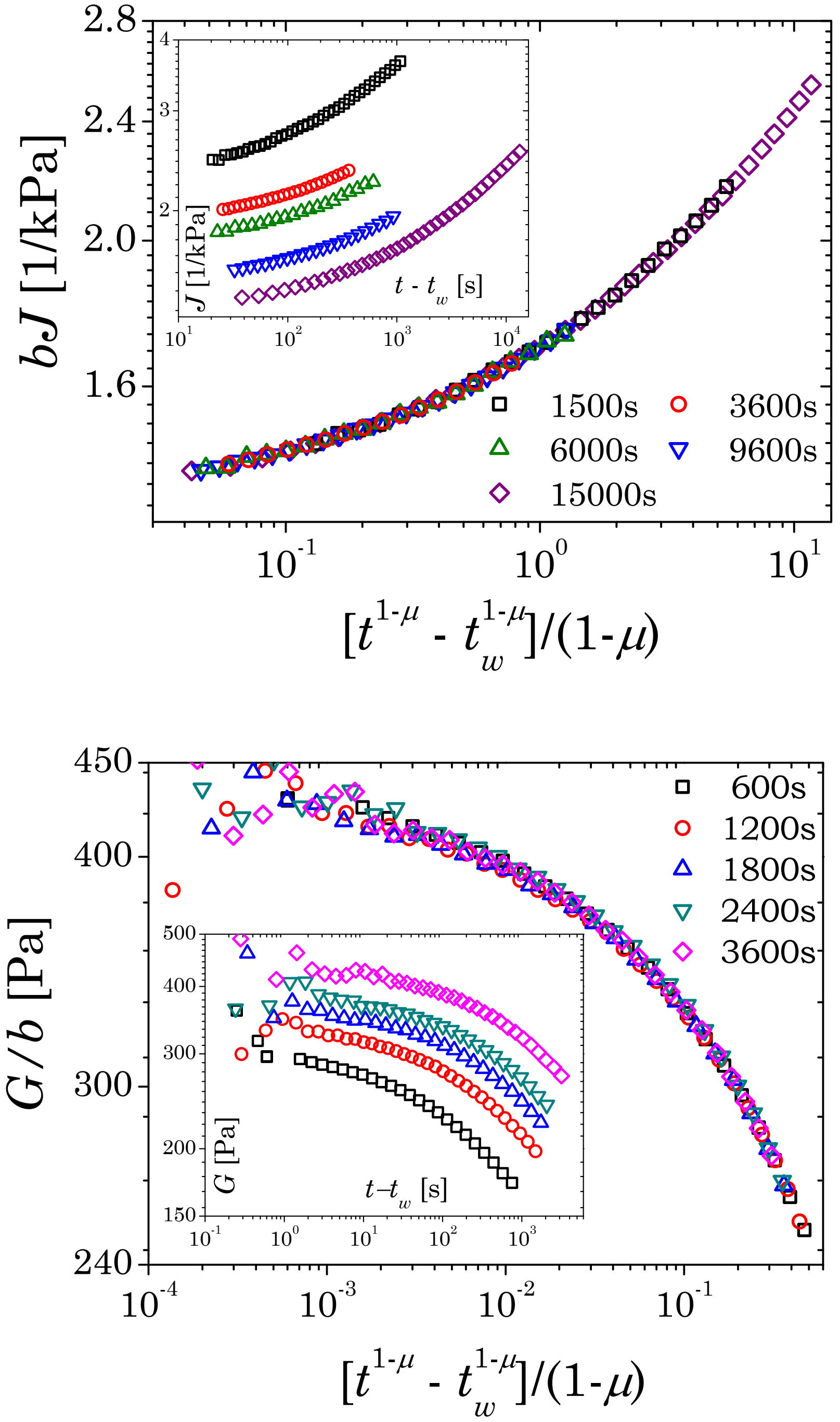

(a)

(b)

**Figure 2.** Time – aging time superposition for creep compliance (a), and for stress relaxation modulus (b) in the effective time domain. The insets in respective figures show evolution of (a) compliance (55°C) and (b) stress relaxation modulus (25°) as a function of time. The vertical shift factor is given by $b=G(t_w)/G(t_{wR})$ in the limit $t-t_w \to 0$. $G(t_{wR})$ is modulus at the reference aging time $(t_{wR})$.

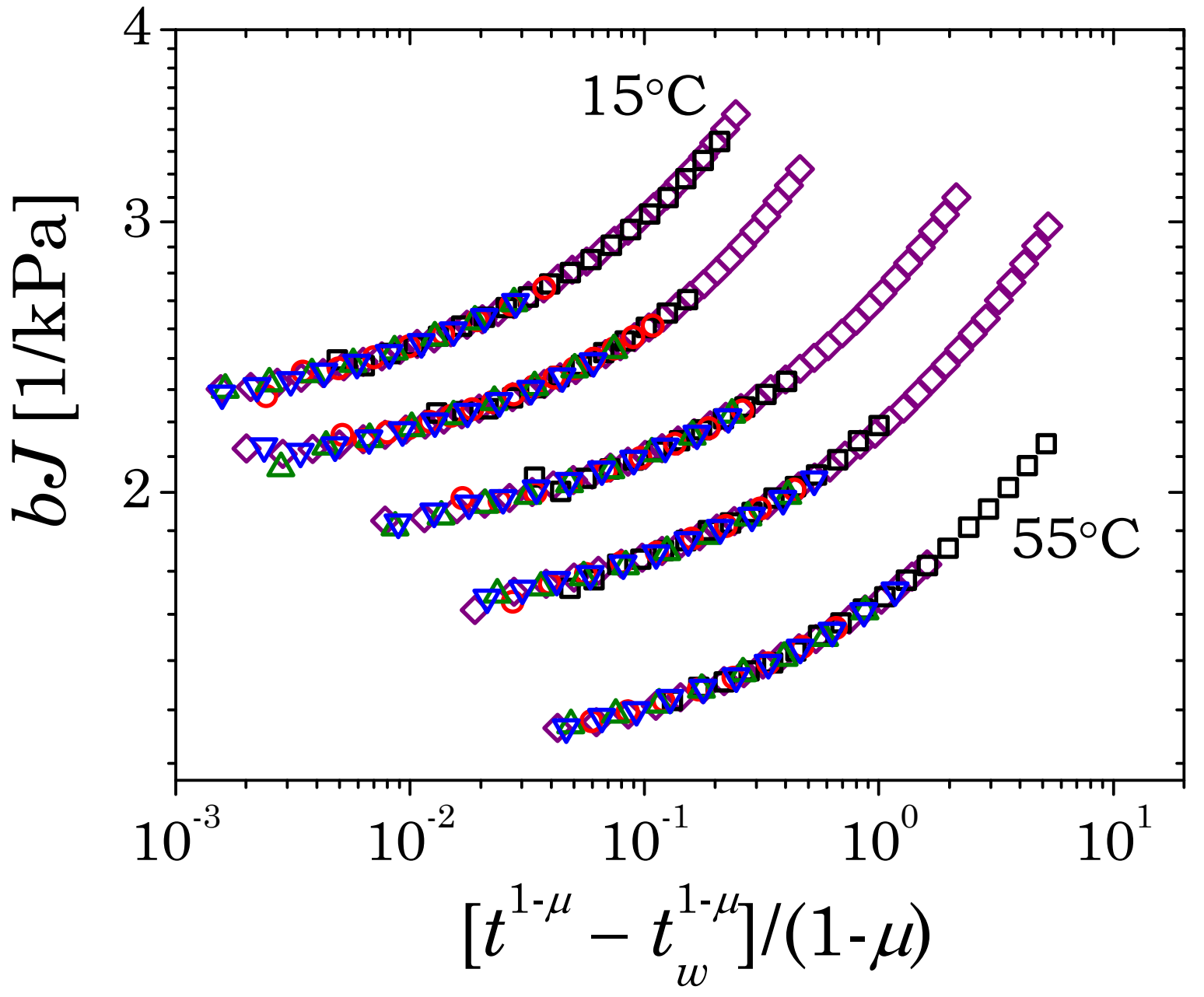

(a)

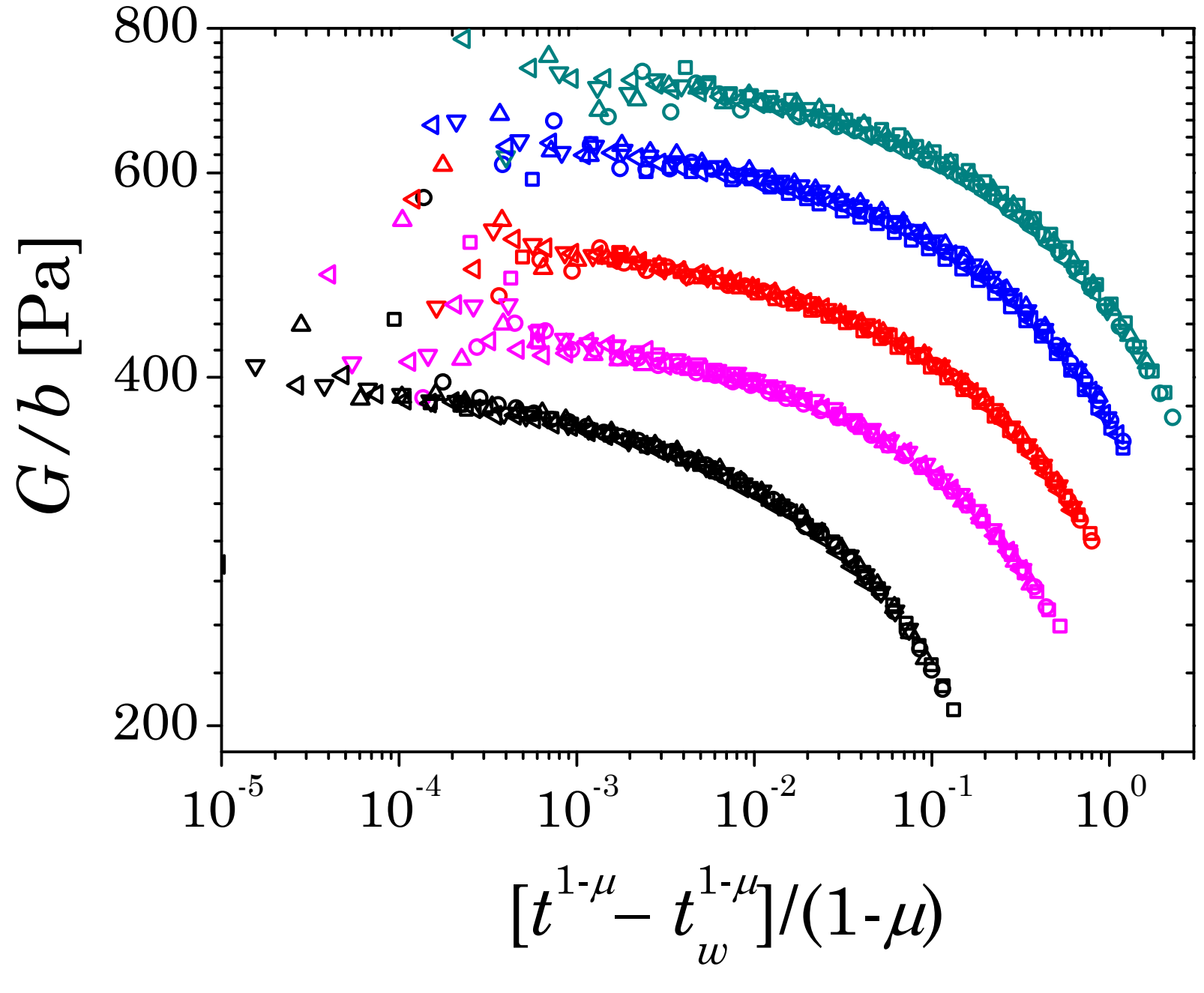

(b)

**Figure 3.** Time – aging time superpositions at different temperatures. (a) For Creep experiments, from top to bottom: 15, 25, 35, 45, 55°C. (b) For stress relaxation experiments, from top to bottom: 55, 45, 35, 25, 15°C.

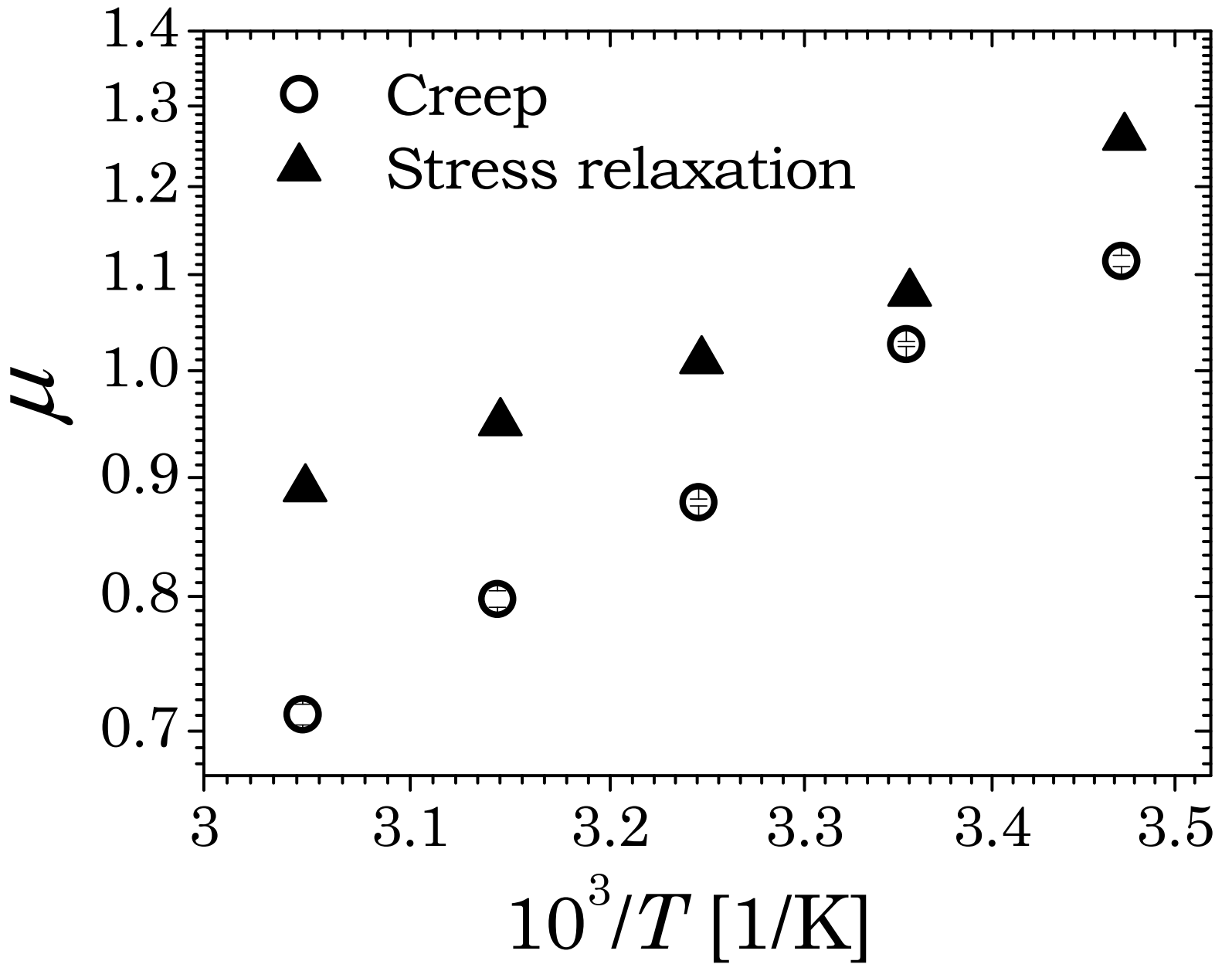

**Figure 4.** Dependence of factor $\mu$ on temperature.

We perform creep and stress relaxation experiments at different aging times over a range of temperatures. In figure 3a and 3b, we plot time – aging time superpositions for creep compliance and relaxation modulus obtained at various temperatures. Due to enhanced elastic modulus and viscosity, compliance induced in the material is lesser for experiments carried out at higher temperatures. Equivalently, relaxation of stress is slower for experiments performed at higher temperatures. In figure 4, we plot value of $\mu$, needed to obtain the time – aging time superposition, as a function of reciprocal of temperature. Over a range of temperatures studied in this work, $\mu$ can be seen to be following linear dependence on $1/T$. If we assume two expressions of relaxation times: $\tau = A\tau_m^{1-\mu} t_w^{\mu}$ and $\tau = \tau_m \exp(E/k_B T)$, to be equivalent and substitute two relations for elastic modulus: $G' = G_0 \ln(t_w/\tau_m)$ and $G' = E/b^3$, we get: $\mu \propto G_0 b^3/k_B T$ (Actually $\mu = \left[G_0 b^3/k_B T\right] - \left[G_0 b^3 \ln A/E\right]$, however the very fact that $\mu$ remains constant with respect to the aging time suggests that the second term is negligible compared to the first term, $E >> k_B T \ln A$). Interestingly experimental observation shown in figure 4 is in qualitative agreement with this dependence.

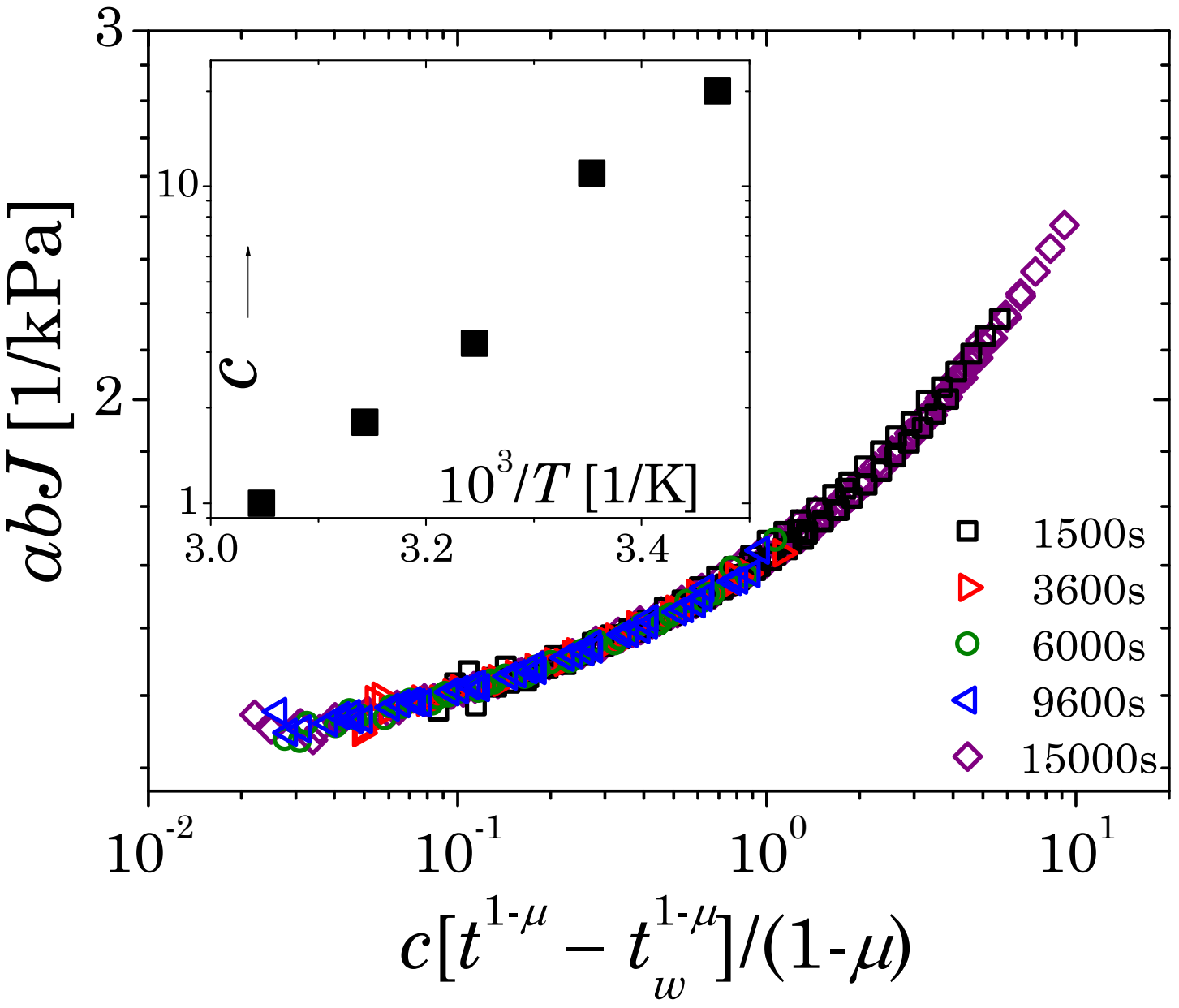

(a)

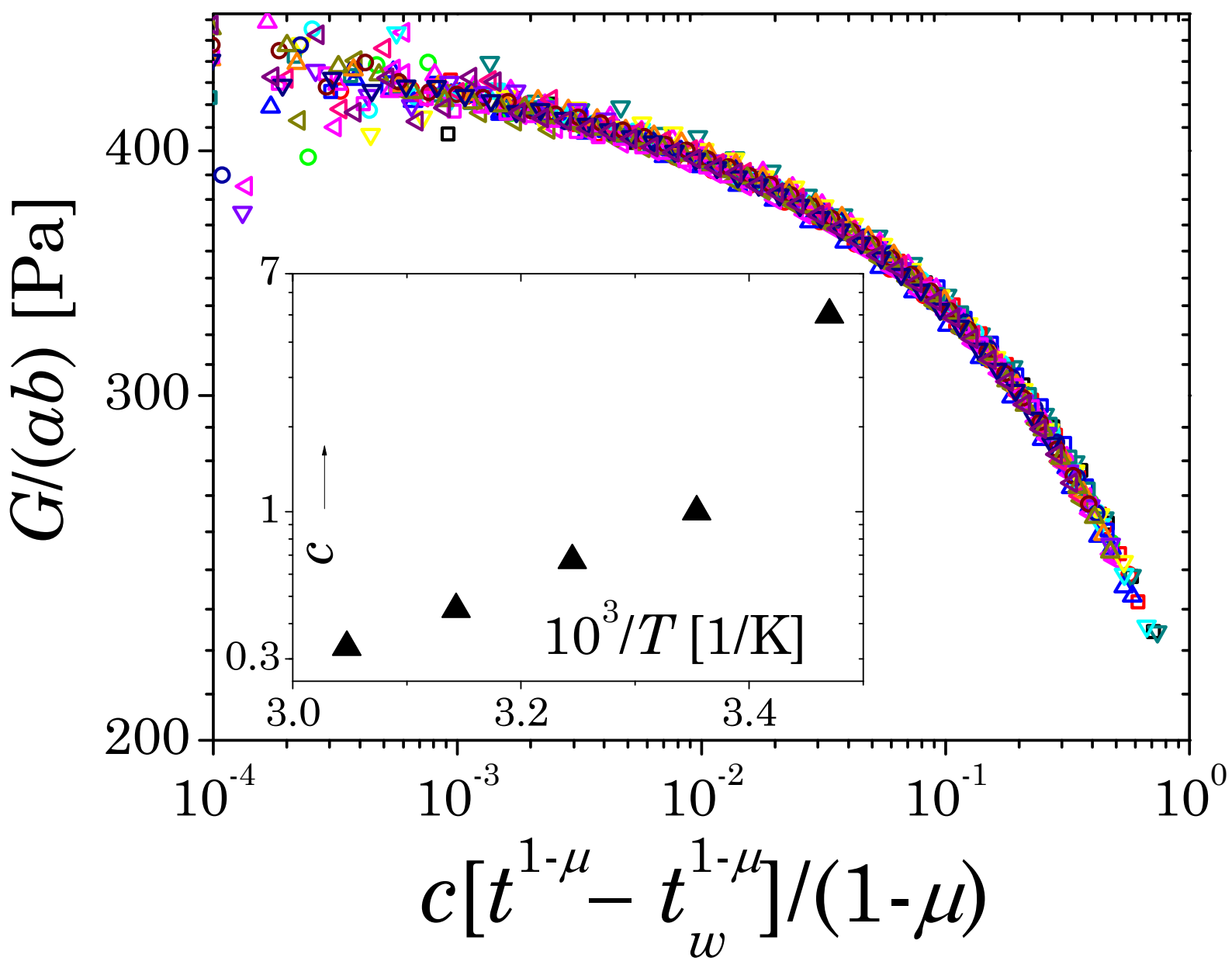

(b)

**Figure 5.** Time – aging time – temperature superposition in an effective time domain for (a) Creep experiments, and (b) Step strain experiments. The insets in respective figures show dependence of horizontal shift factor on temperature. Vertical shift factor is given by $a = G(t_{wR}, T)/G(t_{wR}, T_R)$ in the limit $t - t_w \to 0$. We consider reference temperature ($T_R$) to be 55°C for creep while 25°C for step strain experiments.

It can be seen that the time – aging time superpositions obtained at different temperatures described in figure 3a and 3b have self-similar curvatures. By carrying out horizontal and vertical shifting we obtain comprehensive time – aging time – temperature superposition for creep compliance (figure 5a) and stress relaxation modulus (figure 5b). It should be noted that, in the superposition of creep data we have considered the data obtained at 55°C and $t_w$=15000 s only up to 1500 s of creep time. Motivation behind consideration of partial data is to enable prediction of the remaining data as discussed below. Vertical shift factor is necessary to compensate for enhancement in elastic modulus as a function of temperature at given aging time. On the other hand, horizontal shift factor is necessary to accommodate change in relaxation time as a function of temperature. It should be noted that in an effective time domain the rheological data should be represented as a function of $\left[\xi(t) - \xi(t_w)\right]/\tau_0$ for every temperature. However for convenience, in figure 2 and 3, we divide the difference in effective time by $\tau_0 \tau_m^{\mu-1}/A$ leading to:

$\left\{\left[\xi(t)-\xi(t_w)\right]\big/\left(\tau_0\tau_m^{\mu-1}/A\right)=\left(t^{1-\mu}-t_w^{1-\mu}\right)\big/(1-\mu)\right\}$. Consequently in order to represent abscissa of figure 5 as $\left[\xi(t)-\xi(t_w)\right]\big/\tau_0$ horizontal shift factor should scale as $c\propto\tau_m^{\mu-1}/A$. In order to shift individual superpositions on to a superposition at reference temperature, horizontal shift factor should therefore scale as: $c=\tau_m^{\mu-1}\big/\tau_{mR}^{\mu_R-1}$. As discussed before, if we assume microscopic time to have Arrhenius dependence on temperature $\tau_m=\tau_{m0}\exp\left(\bar{U}/k_BT\right)$, where $\tau_{m0}$ is an attempt time while $\bar{U}$ is energy barrier associated with microscopic movement of the entity within the cage; we get: $\ln c\propto\frac{G_0b^3}{k_BT}\left(\ln(\tau_{m0})+\frac{\bar{U}}{k_BT}\right)$ ($\bar{U}$ should not be confused with $E$, which is barrier height associated with energy well). As shown in the insets of figure 5a and 5b, observed linear dependence of $\ln c$ on reciprocal of temperature suggests increase in relaxation time as a function of temperature.

Time – aging time – temperature superpositions shown in figure 5(a) and 5(b) suggest that the material will follow evolution of creep compliance or stress relaxation modulus in an effective time domain if experiments were carried out at respective reference temperatures. Since at reference temperature, $c=1$; superpositions described in figures 5a and 5b can be transformed back into real time domain as suggested by Shahin and Joshi.[14] If we consider abscissa of figure 5 to be $\theta=\left[t^{1-\mu}-t_w^{1-\mu}\right]\big/(1-\mu)$, real time elapsed since application of deformation field can simply be represented by:

$$t-t_w=\left\{\theta(1-\mu)+t_w^{1-\mu}\right\}^{1/(1-\mu)}-t_w. \qquad (3)$$

As shown in figure 5, superposition extends beyond the original creep or stress relaxation data available at reference temperature and aging time. Therefore, transformation of the superposition from effective time domain to real time domain leads to prediction of long time as well as short time behavior. In figure 6a and 6b, we plot creep and stress relaxation time data at reference temperature and aging times, and compare the same with corresponding prediction by transforming the superposition using eq. (3). For the creep data associated with 55°C and $t_w$=15000 s, data that was not considered in the superposition (1500 to 15000 s) is represented by down triangles. It can be seen that the proposed procedure predicts the overall creep data very well. The benefit of time – aging time – temperature superposition described in figure 5 over time – aging time superposition plotted in figure 2 can be appreciated better from the superposition associated with stress relaxation data wherein we consider compete data in the superposition. The information available from figure 3b is over the range of $\theta$ from $5\times10^{-4}$ to 0.45 while the same from figure 5b is available in the range: $2\times10^{-4}$ to 0.8, which according to eq. (3) will lead to prediction over broader creep times. The procedure also leads to prediction of very short time rheological behavior as well. We feel that prediction of short time behavior is also very important as very small time regimes are usually not accessible due to instrument inertia and data acquisition speeds.

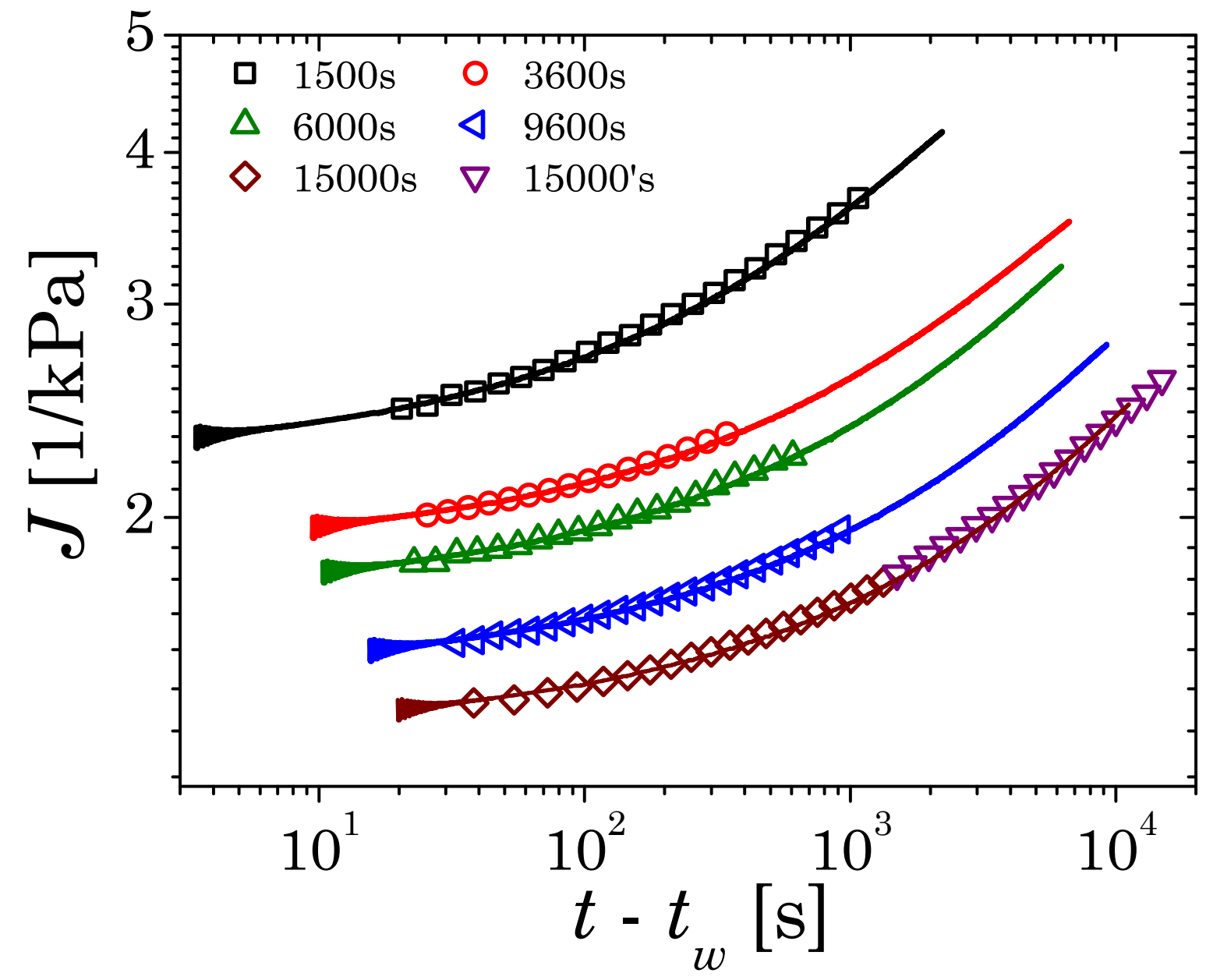

(a)

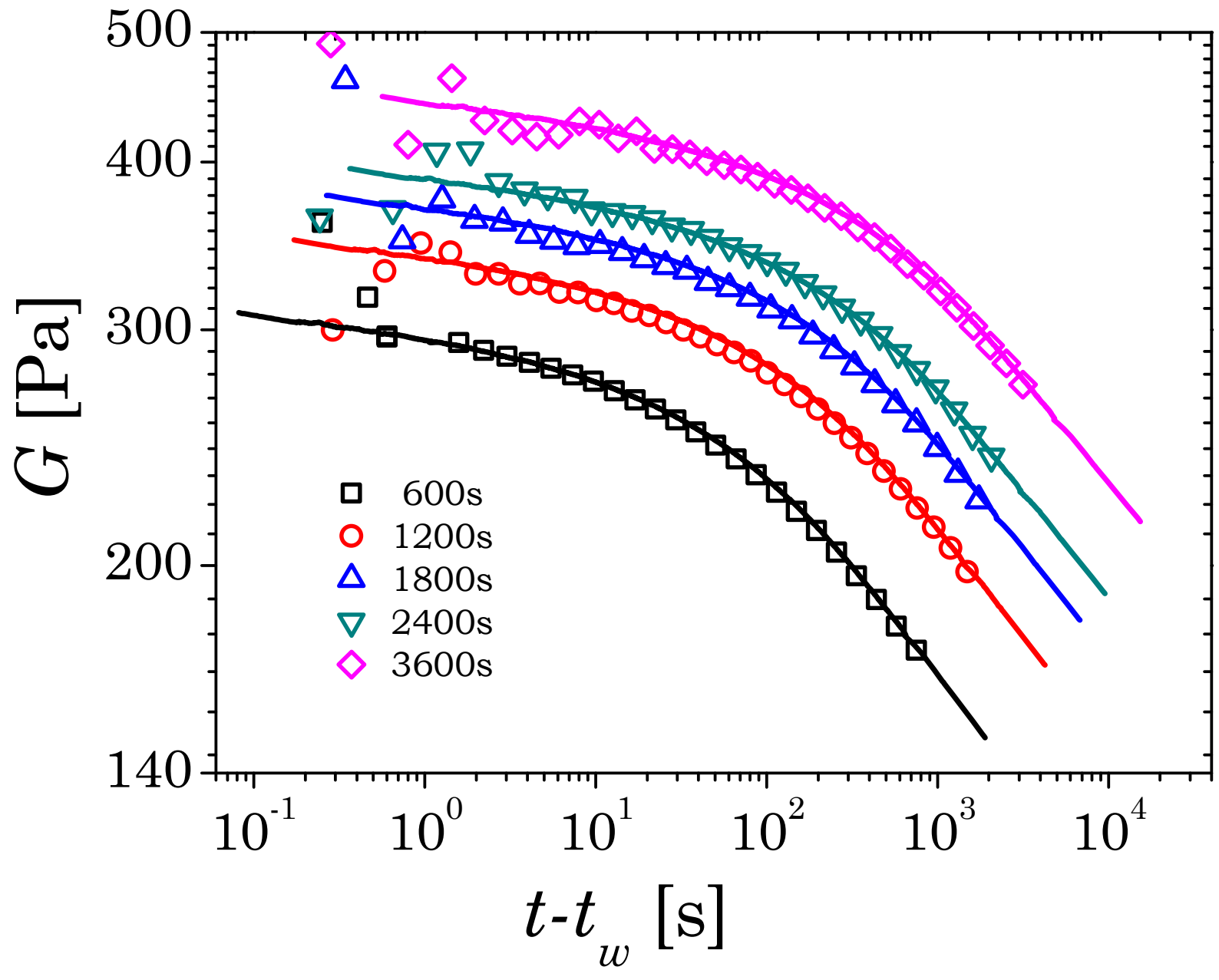

(b)

**Figure 6.** Prediction of the long and short time rheological behavior (a) creep experiments (55°C), and (b) stress relaxation experiments (25°C). Symbols are experimental data while lines are the predictions.

It is important to note that the fundamental basis of time – temperature superposition described in figure 5 and that carried out in equilibrium soft materials such as polymeric liquids is identical. In the later systems relaxation time of the material decreases with increase in temperature and therefore experiments carried out at higher temperatures facilitate description of long time creep or stress relaxation behaviors.[15] For the present system, closer inspection of figures 3 and 5 indicates that superpositions obtained at low temperatures facilitate prediction of long time rheological behavior of experiments carried out at higher temperatures. Conversely, superposition corresponding to high temperatures leads to description of very short time rheological behavior. This observation suggests that at low temperatures aqueous suspension of Laponite has smaller relaxation time at given aging time. Interestingly, figure 4 indicates that $\mu$, which describes logarithmic dependence of relaxation time on aging time $\left(d\ln\tau/d\ln t_w\right)$, decreases with increase in temperature. Considering both these aspects, in figure 7 we schematically describe dependence of relaxation time on aging time as a function of temperature. Figure 7 suggests that, although greater in magnitude, relaxation time at higher temperature evolves slowly.

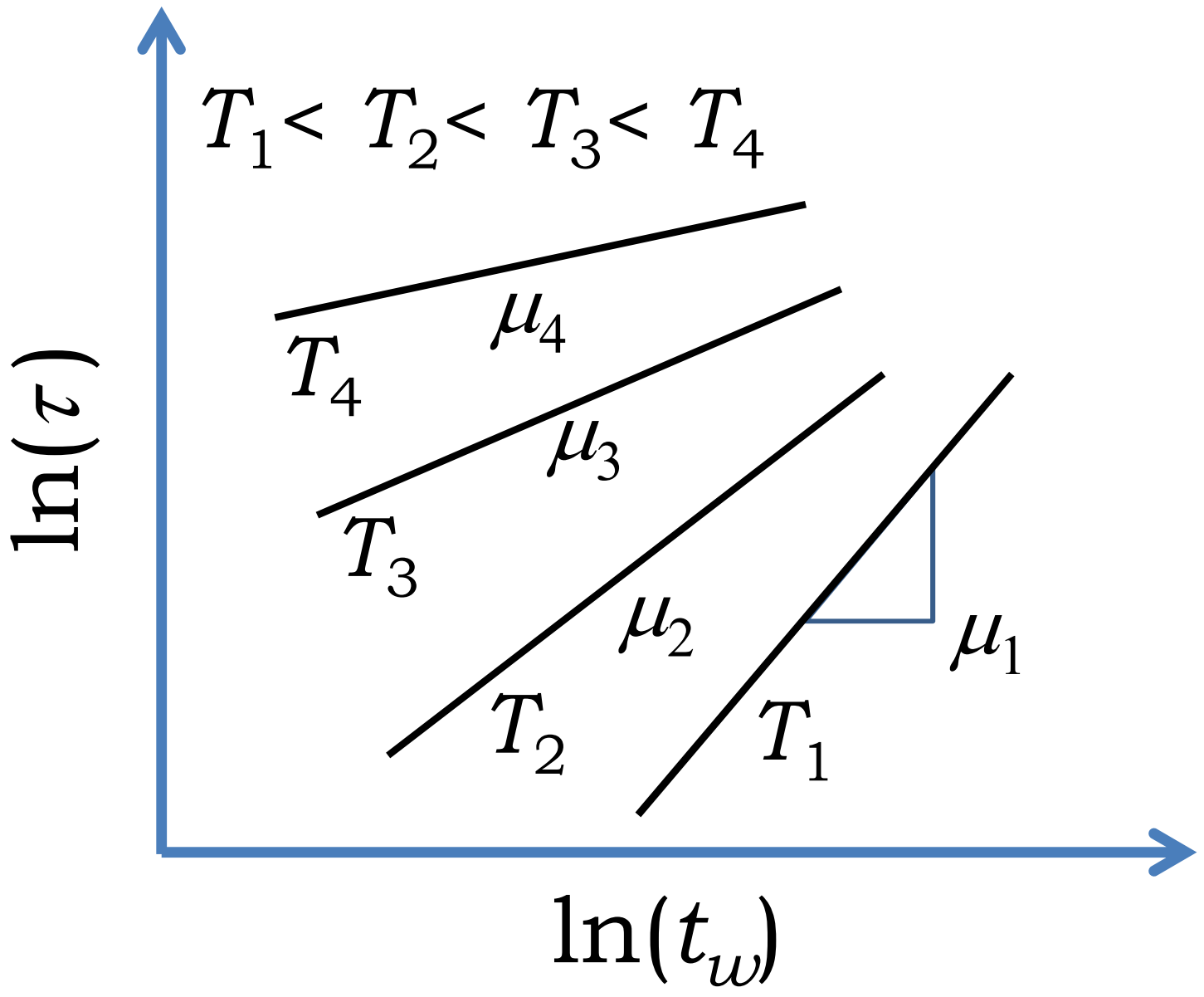

**Figure 7.** Schematic describing dependence of characteristic relaxation time on aging time as a function of temperature.

In order to observe time – temperature superposition for equilibrium soft materials, the shape of relaxation time distribution needs to remain unaffected by change in temperature. Same is also true for soft glassy materials.[13] Therefore, in order to observe time – aging time – temperature superposition, the shape of relaxation time distribution needs to be unaffected by variation of aging time as well as temperature. Similar curvatures of rheological data reported in the insets of figures 2 and that of superpositions shown in figure 3 essentially manifest that the shape of the distribution has remained unaffected by aging time and temperature respectively. We believe that this is the only constraint which needs to be fulfilled for observation of time – temperature superposition for any soft glassy material in the effective time domain. The specific details of how relaxation time and modulus behave as a function of temperature may be material specific and will only affect the predictive capacity of the superposition. We hope that time – temperature superposition procedure developed in this work for soft glassy materials will be tested for variety of non-ergodic systems and will prove helpful in characterizing these industrially important complex fluids.

**IV. Conclusions:**

Soft glassy materials exhibit temporal evolution of their structure and viscoelastic properties. Particularly the relaxation dynamics of soft glassy materials is observed to become sluggish with respect to time. Consequently compliance or modulus exhibits an additional dependence on time at which deformation field was applied disregarding Boltzmann superposition principle in its conventional form. Transformation of the rheological data to effective time domain, wherein all the relaxation processes are rescaled by considering a constant relaxation time, allows application of Boltzmann superposition principle. Compliance or stress relaxation modulus is observed to demonstrate time – aging time superposition when plotted in the effective time domain irrespective of time of application of deformation field. In this work we extend effective time framework to demonstrate time – temperature superposition in creep as well as step strain (stress relaxation) experiments. We employ aqueous Laponite suspension as a model soft glassy material. In this system we observe that relaxation time of the material is greater while it's temporal evolution in logarithmic scale ($d\ln\tau/d\ln t_w$) is weaker at higher temperatures. The time – aging time superpositions obtained at different temperatures show similar curvature in the effective time domain, thereby enabling time – aging time – temperature superposition by horizontally shifting the data. Self-similar curvatures of the creep data suggest that temperature affects only the average value of relaxation time and not the shape of the distribution, which in turn is a criterion for observing time – temperature superposition. Owing to enhanced relaxation time with increase in temperature, experiments carried out over accessible timescales at lower temperatures facilitates prediction of long time behavior at higher temperatures through time – temperature superposition. Similarly prediction of short time data is aided by experiments performed at higher temperatures.

**Acknowledgement:** This work was supported by Department of Science and Technology, Government of India under IRHPA scheme.

**References:**

1. L. Cipelletti and L. Ramos, *J. Phys. Cond. Mat.*, 2005, **17**, R253–R285.
2. P. Coussot, *Lecture Notes in Physics*, 2006, **688**, 69-90.
3. M. E. Cates and M. R. Evans, eds., *Soft and fragile matter*, The institute of physics publishing, London, 2000.
4. G. B. McKenna, T. Narita and F. Lequeux, *Journal of Rheology*, 2009, **53**, 489-516.
5. S. A. Rogers, P. T. Callaghan, G. Petekidis and D. Vlassopoulos, *Journal of Rheology*, 2010, **54**, 133-158.
6. D. J. Wales, *Energy Landscapes*, Cambridge University Press, Cambridge, 2003.
7. C. Christopoulou, G. Petekidis, B. Erwin, M. Cloitre and D. Vlassopoulos, *Philosophical Transactions of the Royal Society A: Mathematical, Physical and Engineering Sciences*, 2009, **367**, 5051-5071.
8. N. Koumakis and G. Petekidis, *Soft Matter*, 2011, **7**, 2456-2470.
9. P. Coussot, *Rheometry of Pastes, Suspensions and Granular Materials-Application in Industry and Environment*, Wiley, Hoboken, 2005.
10. J. Mewis and N. J. Wagner, *Advances in Colloid and Interface Science*, **147-148**, 214-227.
11. S. M. Fielding, P. Sollich and M. E. Cates, *Journal of Rheology*, 2000, **44**, 323-369.
12. I. L. Hopkins, *J. Polym. Sci.*, 1958, **28**, 631-633.
13. B. Baldewa and Y. M. Joshi, *Soft Matter, DOI:10.1039/C1SM06365K.*, 2011.
14. A. Shahin and Y. M. Joshi, *Phys. Rev. Lett.*, 2011, **106**, 038302.
15. R. B. Bird, R. C. Armstrong and O. Hassager, *Dynamics of Polymeric Liquids, Fluid Mechanics*, Wiley-Interscience, New York, 1987.

16. H. A. Barnes, J. F. Hutton and K. Walters, *An Introduction to Rheology*, Elsevier, Amsterdam, 1989.
17. L. C. E. Struik, *Physical Aging in Amorphous Polymers and Other Materials*, Elsevier, Houston, 1978.
18. S. M. Fielding, Ph.D. Thesis, University of Edinburgh, 2000.
19. P. A. O'Connell and G. B. McKenna, *Polym. Eng. Sci.*, 1997, **37**, 1485-1495.
20. G. F. Rodriguez, G. G. Kenning and R. Orbach, *Physical Review Letters*, 2003, **91**, 037203.
21. P. Sibani and G. G. Kenning, *Physical Review E*, 2010, **81**, 011108
22. M. Cloitre, R. Borrega and L. Leibler, *Phys. Rev. Lett.*, 2000, **85**, 4819-4822.
23. P. Coussot, H. Tabuteau, X. Chateau, L. Tocquer and G. Ovarlez, *J. Rheol.*, 2006, **50**, 975-994.
24. V. Awasthi and Y. M. Joshi, *Soft Matter*, 2009, **5**, 4991–4996.
25. Y. M. Joshi and G. R. K. Reddy, *Phys. Rev. E*, 2008, **77**, 021501-021504.
26. A. Shaukat, A. Sharma and Y. M. Joshi, *Rheologica Acta*, 2010, **49**, 1093-1101.
27. C. Derec, G. Ducouret, A. Ajdari and F. Lequeux, *Physical Review E*, 2003, **67**, 061403.
28. B. Baldewa and Y. M. Joshi, *Polymer Engineering and Science*, 2011, **51**, 2084-2091.
29. D. Bonn, S. Tanasc, B. Abou, H. Tanaka and J. Meunier, *Phys. Rev. Lett.*, 2002, **89**, 157011-157014.
30. http://www.laponite.com.
31. A. Shahin and Y. M. Joshi, *Langmuir*, 2010, **26**, 4219–4225.
32. R. A. L. Jones, *Soft Condensed Matter*, Oxford University Press, Oxford, 2002.
33. A. J. Kovacs, *J Polym. Sci.*, 1958, **30**, 131-147.
34. G. Ovarlez and P. Coussot, *Phys. Rev. E*, 2007, **76**, 011406.
35. P. Agarwal, H. Qi and L. A. Archer, *Nano Letters*, 2009, **10**, 111-115.
36. X. Di, K. Z. Win, G. B. McKenna, T. Narita, F. Lequeux, S. R. Pullela and Z. Cheng, *Physical Review Letters*, 2011, **106**, 095701.
37. P. Agarwal, S. Srivastava and L. A. Archer, *Phys. Rev. Lett. in press.*, 2011.